# Soft x-rays absorption and high-resolution powder x-ray diffraction study of superconducting $Ca_xLa_{1-x}Ba_{1.75-x}La_{0.25+x}Cu_3O_y$ system

S. Agrestini[1], S. Sanna[2,*], K. Zheng[3], R. De Renzi[3], E. Pusceddu[4], G. Concas[4], N. L. Saini[5], and A. Bianconi[6]

[1] *Max-Planck-Institut für Chemische Physik fester Stoffe Nöthnitzer Str. 40 D-01187 Dresden, Germany*

[2] *Dipartimento di Fisica and CNISM, Universit`a di Pavia, viale Bassi 6, 27100 Pavia, Italy*

[3] *Dipartimento di Fisica and CNISM, Università di Parma, viale Usberti 7A, 43100 Parma, Italy*

[4] *Dipartimento di Fisica and INSTM, Università di Cagliari, S.P. Monserrato-Sestu km 0.700, I-09042 Monserrato (CA), Italy*

[5] *Dipartimento di Fisica, Sapienza University of Rome, P. le Aldo Moro 2, 00185 Roma, Italy*

[6] *Rome International Centre for Materials Science, Superstripes Via dei Sabelli 119A, 00185 Roma, Italy*

*Corresponding author. E-mail address: samuele.sanna@unipv.it

**Abstract.** We have studied the electronic structure of unoccupied states measured by O K-edge and Cu L-edge x-ray absorption spectroscopy (XAS), combined with crystal structure studied by high resolution powder x-ray diffraction (HRPXRD), of charge-compensated layered superconducting $Ca_xLa_{1-x}Ba_{1.75-x}La_{0.25+x}Cu_3O_y$ ($0 \leq x \leq 0.4$, $6.4 \leq y \leq 7.3$) cuprate. A detailed analysis shows that, apart from hole doping, chemical pressure on the electronically active $CuO_2$ plane due to the lattice mismatch with the spacer layers greatly influences the superconducting properties of this system. The results suggest chemical pressure to be the most plausible parameter to control the maximum critical temperatures ($T_c^{max}$) in different cuprate families at optimum hole density.

***Keywords:*** chemical pressure, hole content, x-ray absorption spectroscopy, cuprate, high $T_c$ superconductivity

***PACS:*** 78.70.Dm, 61.05.cp, 74.25.Dw, 74.25.Jb, 74.72.Gh

**Introduction**

For more than two decades, the phase diagram of superconductivity in layered cuprates has been described by the universal dome $T_c(h)$ where the transition temperature ($T_c$) is controlled by charge density (h) in the $CuO_2$ plane, measured by the number of holes per Cu ion. The insulator-to-superconductor transition occurs at h~0.06 and the maximum $T_c^{max}$ occurs at an optimum hole density h~0.16. However, the key problem is that $T_c^{max}$ varies from a few tens of Kelvin to 135 K in different cuprate families [1-3]. This has raised the importance of the structural topology for controlling the $T_c^{max}$ in these materials. Indeed, recently it has been argued that electronic inhomogeneity controlled by organization of defects, lattice distortions, interface states, nanoscale phase separation in the layered oxides has substantial influence in the superconducting properties of cuprates [4-9]. This appears to be common to all layered superconductors as diborides [10-13] and more recently pnictides [14-19].

Here it should be specified that large structural differences from one cuprate family to another one hinders the investigation of what drives the $T_c^{max}$ in these materials. Furthermore, it is usually difficult to separate the specific contributions of the charge density and the hosting lattice, since both of them are affected simultaneously by the doping in the system, either by chemical substitution or oxygen loading. The charge-compensated cuprate $Ca_xLa_{1-x}Ba_{1.75-x}La_{0.25+x}Cu_3O_y$ (CLBLCO) can be a special test case to overcome the problem, since the $T_c$ can be changed along the dome profile from the underdoped to the overdoped regime by varying y (hole density), and the $T_c^{max}$ at optimum doping ($y^{opt}$ ~7.13) can be changed by varying x, with the $T_c^{max}$ changing from about 45K to 81K (Fig. 1). Here we have exploited combination of soft X-ray absorption spectroscopy (XAS) and high resolution powder x-ray diffraction (HRPXRD) to study both the electronic structure of the unoccupied states (hole states) and the crystal structure, addressing hole density, disorder and the misfit strain between different lattice parameters of different layers effects on superconductivity of CLBLCO as a function of oxygen (hole doping) and calcium (structure and disorder) contents. The analysis of O K- and Cu L-edge XAS reveals that the hole density in the CLBLCO at optimum doping (y≈7.13), remains constant in spite of a significant increase of $T_c^{max}$ from 45 K to 81 K when calcium content is changed from x=0 to 0.4. The Rietveld analysis of the HRPXRD shows a systematic linear dependence of $T_c^{max}$ on the in-plane lattice parameter. The results suggest that in these heterostructures at atomic limit the superlattice misfit strain due to the lattice mismatch between the $CuO_2$ plane and the spacer layer is an important parameter to be considered separately from disorder to unravel the complex behavior of high temperature superconducting cuprates.

The main result of this study on CLBLCO has already been published elsewhere [20]. Here we report the details of the analysis of the whole XAS spectra and we add new results on the effect on structural disorder, by studying the evolution of the shape of selected Bragg peaks of the HRPXRD patterns as a function of Ca and oxygen contents.

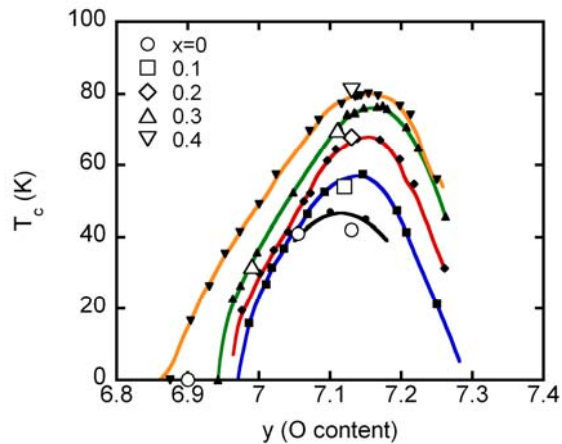

**Fig. 1** The phase diagram of $Ca_xLa_{1-x}Ba_{1.75-x}La_{0.25+x}Cu_3O_y$ (0≤x≤0.4, 6.9≤y≤7.3) from Ref.[21]. The open symbols indicate the CLBLCO samples investigated in this work. The solid lines are guide to the eyes.

## 2. Experimental methods

The CLBLCO polycrystalline samples were prepared by the standard solid-state synthesis method. The oxygen content in the underdoped samples was tuned by a topotactic technique [22] of oxygen exchange between donor (y~7.3) and acceptor (y~6.4) end terms. The optimum doped samples (y≈7.13) were obtained by annealing the samples at 500°C under the oxygen pressure of 30-50 atm. The phase purity and superconducting properties of the samples were characterized before the XAS and HRPXRD measurements by standard x-ray diffraction, electric resistivity and magnetization measurements.

The O K-edge and Cu L-edge XAS was carried out on ISIS beamline of BESSY storage ring (Berlin, Germany), with energy resolutions of 220 meV at 530 eV, and 520 meV at 930 eV. The samples were scraped in situ with a diamond file to get a fresh surface before the XAS measurements under a base pressure lower than $5\times10^{-10}$ mbar. Non-surface-sensitive fluorescence yield was collected with a windowless high purity Ge-detector. The fluorescence yield data were corrected for the incident photon flux, and were normalized in an range of ~70 eV above the threshold energy. The photocurrent yield was collected as well, to check the consistency of the self-absorption correction. The high resolution powder x-ray diffraction measurements on the same samples were performed with the wavelength λ= 0.4 Å on the ID31 beamline at ESRF synchrotron in Grenoble.

## 3. Results and discussion

*3.1 O K-edge X-ray absorption spectra*

Fig. 2 shows typical O K-edge XAS measured on CLBLCO with different oxygen and Ca contents. The XAS spectra are quite similar to those of Y123 [23,24], and a similar interpretation of the spectral features can be used also for the present case. The region of the XAS spectra between 527-530 eV (the so-called pre-edge region) probes the unoccupied O 2p electronic states near the Fermi level, reached by direct 1s → 2p transition. Therefore, the pre-edge peak can be used to have direct information on the hole density [23-25]. In this energy region, there are two main features (inset of Fig.2): a peak at 528.3 eV due to transitions from O 1s level to the valence band (VB) of mainly O 2p character, and a second peak at 529.3 eV, associated to transitions from O 1s level to upper Hubbard band (UHB), formed by the hybridization of $Cu3d^9$ and $Cu3d^{10}L$ states (where L is a ligand hole). As the oxygen content is increased from y=6.4 to y=7.13, the two spectral features display an opposite behaviour: the peak at 528.3 eV gains spectral weight while the one at 529.3 loses. The spectral-weight transfer from UHB to VB is similar to the one observed in Y123 [23,24] and is due to the creation of hole states in the VB by oxygen doping of the compound. On the contrary one can hardly see any difference between the spectra of the samples x=0 (y=7.13) and x=0.4 (y=7.12) in the pre-edge region, which indicates that Ca substitution is isoelectronic, i.e. it does not change the hole concentration in CLBLCO.

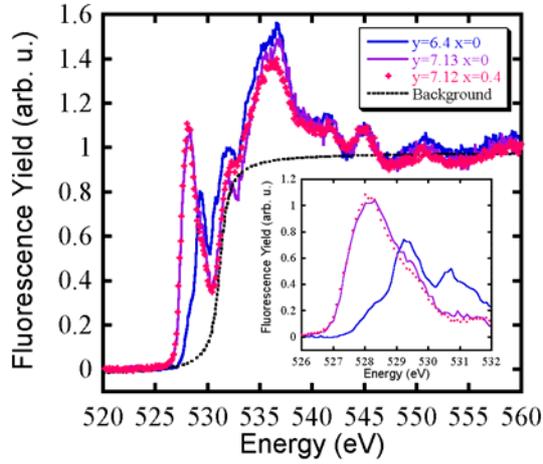

**Fig. 2** O K edge XAS spectra of $Ca_xLa_{1-x}Ba_{1.75-x}La_{0.25+x}Cu_3O_y$ for different samples measured at 20 K. The inset shows the pre-edge region of the spectra after subtraction of the atomic absorption, approximated by an arctan function (dashed curve).

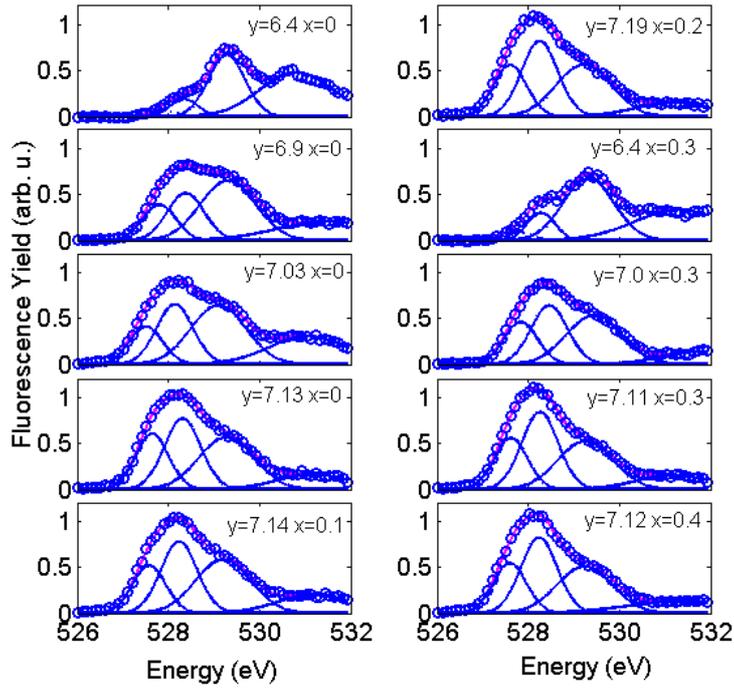

**Fig. 3** Profile fitting of the O K edge XAS spectra of the $Ca_xLa_{1-x}Ba_{1.75-x}La_{0.25+x}Cu_3O_y$ compounds with $6.4 \leq y \leq 7.13$, $0 \leq x \leq 0.4$: experimental data (empty circles) and sum of four Gaussian components (solid curve). The dash-dotted lines indicate the Gaussian functions describing the contribution due to: O 2p hole states in CuO chains (peak at ~527.6 eV); ZR states (peak at ~528.3 eV); UHB (peak at ~529.3 eV).

In order to obtain information on the site-specific distribution of the holes, after the subtraction of the atomic absorption approximated by an inverse tangent function (dash line in Fig. 2), the O K-edge spectra in the range of 526-532 eV were analyzed by profile fitting using a sum of Gaussian functions (see Fig. 3). Each spectrum was fitted with 3 Gaussian functions centered at ~527.6, 528.3 and 529.3 eV, respectively, and a Gaussian component from other signals above 530 eV (see Fig. 3). In the fitting procedure the profile parameters (height, width and position) of each Gaussian function were iterated. The energy positions were found to be identical for all samples within the experimental uncertainties. The full width at half maximum (FWHM) of each Gaussian component is different: 0.70 eV for the peak at lower energy and 1.00 eV for the other two peaks. It is possible to identify these Gaussian peaks as different electronic transitions from the O 1s initial states, consistent with the peak assignment employed by Nücker et al. [23,24] for the XAS spectrum of Y123. The peak at ~527.6 eV is attributed to the transition in the O 2p hole states due to the CuO chains (CH). The peak at 528.3 eV has a contribution from in-plane O 2p hole states hybridized with Cu $3d_{x2-y2}$ orbital. The peak at 529.5 eV represents transition to the upper Hubbard band (UHB). It is fair to mention that the lower energy peaks are not purely due to Cu-O chains and Zhang Rice (ZR) singlet, but a contribution of the O 2p hole states in the apical oxygen is also mixed. Previous XAS measurements on Y123 [23,24] showed that this last contribution is extended over the entire valence band energy range (527-529 eV). Therefore it is impossible to distinctly identify the said contribution in the present data obtained on polycrystalline samples.

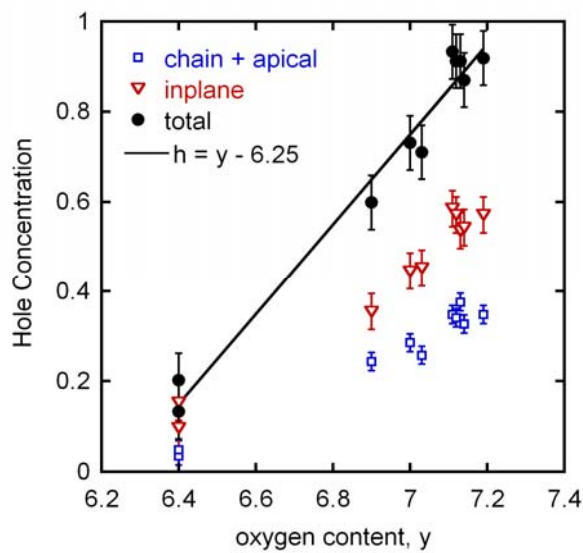

**Fig. 4** The hole concentration, as a function of oxygen content in $Ca_xLa_{1-x}Ba_{1.75-x}La_{0.25+x}Cu_3O_y$, obtained by the integrated areas of the Gaussian components at ~527.6 and ~528.3 eV, plotted together with the expected evolution (solid line) h = y - 6.25.

We could make a qualitative estimation of the holes in the chains, $n_{chain}$, and in the $CuO_2$ planes, $n_{plane}$, from the above analysis of Gaussian components at ~527.6 eV (CH signal) and at ~528.3 eV (ZR signal). Note that the total number of holes in the VB, h=1, corresponding to y=7.25 was used as scaling factor. The evaluated holes on different sites are reported in table 1 and their evolution as a function of oxygen content is plotted in Fig. 4. The total hole concentration in the valence band (black dots) follows nicely the linear expected relation h = y - 6.25 (blue line in the figure). Also the hole numbers $n_{chain}$ and $n_{plane}$ (empty symbols in Fig. 4) increase linearly with oxygen content. On the contrary the concentration and distribution of the doped holes in the chains and in the $CuO_2$ planes remains nearly unchanged in the whole range of Ca concentration 0≤x≤0.4 for CLBLCO samples with the same oxygen content (see TABLE 1). The small increase of $n_{plane}$ by 0.015 holes between the samples y=7.13, x=0 and y=7.12, x=0.4 cannot explain a nearly doubling of $T_c^{max}$ from 45 K to 80 K. Therefore, the present measurements probing the hole distribution rule out the hypothesis of a charge transfer between the chain sites to the $CuO_2$ planes proposed in some previous studies [27,28] to explain $T_c$ behavior with Ca concentration. The results are also consistent with previous bond-valence-sum calculations [29].

For the optimumally doped CLBLCO (y~7.13), the estimated number of holes in the CuO chains is ~0.33, and the number of total holes in $CuO_2$ plane is ~0.55. There are two $CuO_2$ planes in the unit cell, so the holes per Cu site should be ~0.27 at optimum doping. This is higher than that reported in earlier studies [24], where the number of the in-plane holes is 0.2 for $YBa_2Cu_3O_{6.91}$. This over estimation should be due to the contribution of apical oxygen, that can't be separated in the present data. However, by a comparison with single crystal the XAS measurements of Y123 [24] we can deduce that the apical oxygen signal should be contributing by about 30% to the CH and ZR transitions in our XAS spectra. Therefore, we cannot rule out completely the possibility that Ca substitution could cause a charge transfer between the apical oxygen sites and the $CuO_2$ planes. However, the XAS data show that the ratio of the chain to ZR changes only slightly as a function of Ca content. As a consequence, in order to maintain this ratio constant then one should also suppose a simultaneous charge transfer from the apical oxygen sites to both the CuO chains and the $CuO_2$ planes, which has never been observed in the related YBCO system and would be very hard to explain. Furthermore, such a hypothesis would be in contradiction with the previous bond-valence-sum calculations [29].

*3.2 Cu $L_3$-edge X-ray absorption spectra*

As for the O K edge, corresponding XAS measurements and line profile analysis were also performed for the Cu $L_3$-edge. Figure 5 shows representative Cu $L_3$ edge XAS spectra measured on CLBLCO samples. After subtracting the atomic absorption, approximated by an arctan function, each Cu $L_3$ edge XAS spectrum was fitted with two Gaussian components centered at ~931.4 eV, ~932.9 eV (see Fig. 6). For heavily underdoped samples y=6.4, x=0 and y=6.4, x=0.3, a third component at ~934.3 eV is needed. The main $Cu^{2+}$ peak at ~931.4 eV is due to Cu $3d^9 \rightarrow$ Cu $\underline{2p}3d^{10}$ transitions and its intensity is independent of the oxygen and Ca contents. Instead, when oxygen doping increases from heavily underdoped y=6.4 (x=0) to optimum doped y=7.13 (x=0), the feature around 932.9 eV gains its weight, while the feature around 934.3 eV disappears. The shoulder at ~932.9 eV is associated to Cu $3d^9\underline{L} \rightarrow$ Cu $\underline{2p}3d^{10}\underline{L}$ transitions and its spectral weight is proportional to the hole number in the valence band [23,24,30-33]. The peak at ~934.3 eV is present only in the spectra of the heavily underdoped samples (y=6.4, x=0 and y=6.4, x=0.3) and is ascribed to $Cu^{1+}$ states in the O(4)-Cu(1)-O(4) dumbbell [23-26].

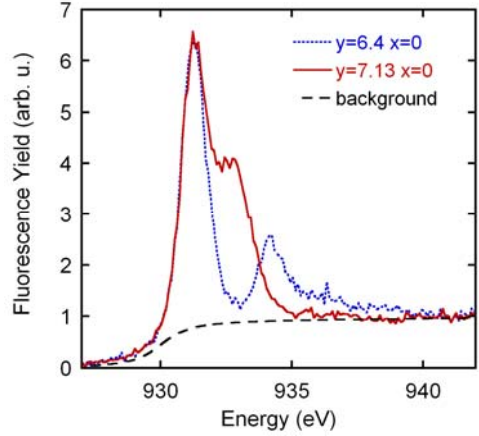

**Fig. 5** Cu $L_3$ edge XAS spectra of $Ca_xLa_{1-x}Ba_{1.75-x}La_{0.25+x}Cu_3O_y$, measured at 20 K, along with the atomic absorption, approximated by an arctan function (dashed curve).

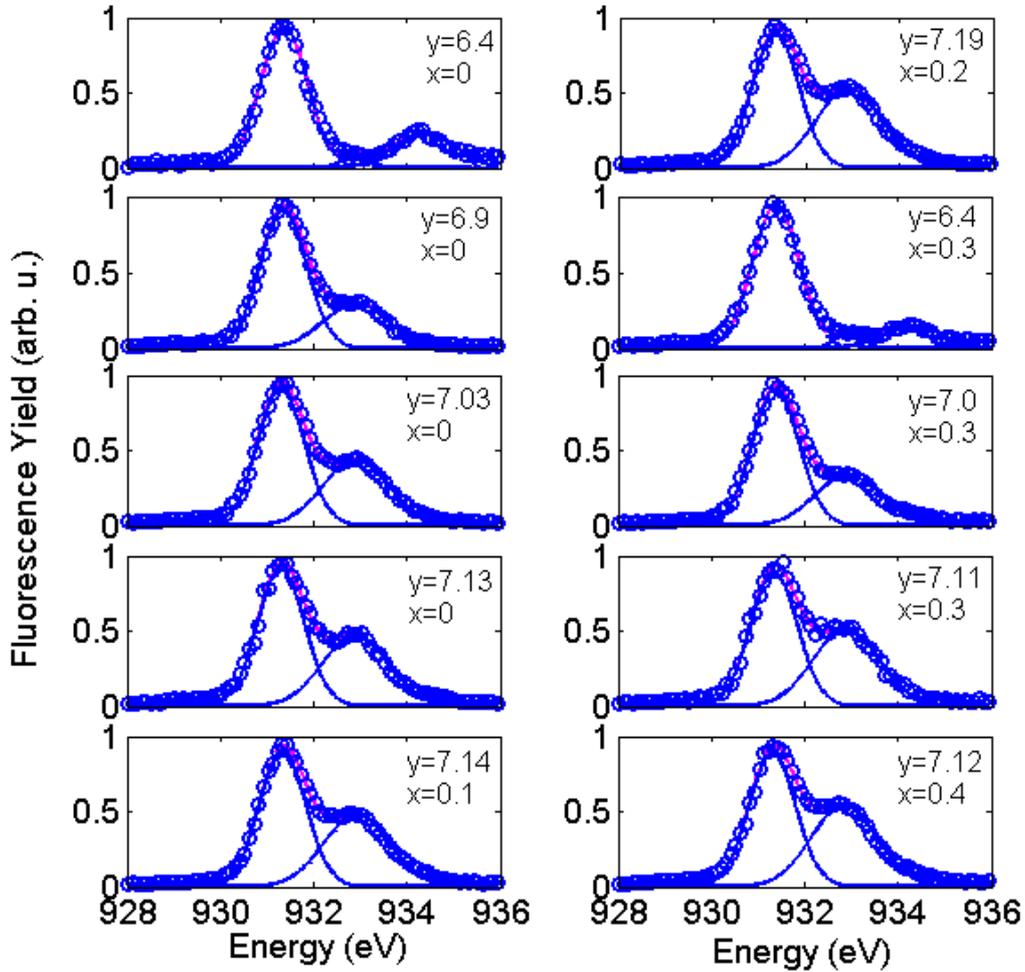

**Fig. 6** Gaussian fitting of the Cu $L_3$ edge XAS of the $Ca_xLa_{1-x}Ba_{1.75-x}La_{0.25+x}Cu_3O_y$ compounds with $6.4 \leq y \leq 7.13$, $0 \leq x \leq 0.4$. The peak at ~931.4 eV is due to Cu $3d^9 \rightarrow$ Cu $\underline{2p}3d^{10}$ transitions while shoulder at ~932.9 eV corresponds to Cu $3d^9\underline{L} \rightarrow$ Cu $\underline{2p}3d^{10}\underline{L}$ transitions ($\underline{L}$ denotes a ligand hole in O 2p state). The peak at ~934.3 eV is from the Cu(1) states in the O(4)-Cu(1)-O(4) dumbbell in the underdoped samples.

| $Ca_xLa_{1-x}Ba_{1.75-x}La_{0.25+x}Cu_3O_y$ | $n_{chain}$ (O K edge) (±0.02) | $2 \times n_{plane}$ (O K edge) (±0.04) | $n_{tot}$ (O K edge) (±0.06) | $n_{tot}$ (Cu L edge) (±0.05) |
|---|---|---|---|---|
| y=6.40 x=0.0 | 0.03 | 0.1 | 0.13 | 0.06 |
| y=6.90 x=0.0 | 0.24 | 0.36 | 0.6 | 0.55 |
| y=7.03 x=0.0 | 0.26 | 0.45 | 0.71 | 0.82 |
| y=7.13 x=0.0 | 0.38 | 0.54 | 0.91 | 0.88 |
| y=7.14 x=0.1 | 0.33 | 0.54 | 0.87 | 0.91 |
| y=7.19 x=0.2 | 0.35 | 0.57 | 0.92 | 0.99 |
| y=6.40 x=0.3 | 0.05 | 0.15 | 0.2 | 0.09 |
| y=7.00 x=0.3 | 0.29 | 0.45 | 0.73 | 0.63 |
| y=7.11 x=0.3 | 0.35 | 0.58 | 0.93 | 0.97 |
| y=7.12 x=0.4 | 0.34 | 0.57 | 0.91 | 1.02 |

**TABLE 1** Hole distribution as determined from the profile analysis of the O K-edge and Cu L-edge XAS. The values are normalized by fixing the total number of the introduced holes to one when y=7.25.

The total holes, obtained from the Cu $L_3$ XAS line shape analysis are also included in TABLE 1, together with the site-specific holes obtained from the O K-edge analysis. The holes estimated from Cu $L_3$-XAS are in agreement with the values estimated from O K-edge XAS. Therefore, the isoelectronic character of the Ca substitution is confirmed by the analysis of Cu $L_3$-edge XAS, but because of the polycrystalline samples we cannot distinguish the contributions from chains and planes as they overlap in the Gaussian component at ~932.9 eV.

*3.3 High-resolution X-ray diffraction*

Our XAS results suggest that Ca substitution does not contribute neither to the total hole doping nor to the hole content in the $CuO_2$ planes. Therefore, other mechanisms should be responsible for the increase of $T_c^{max}$ from 45 K to 81 K with changing Ca content. In the charge-compensated $Ca_xLa_{1-x}Ba_{1.75-x}La_{0.25+x}Cu_3O_y$ compound, Ca atoms partially substitute La atoms and "push" the substituted La atoms onto Ba sites. Because of the very different ionic radii of $Ca^{2+}$ (112 pm), $La^{3+}$ (116 pm) and $Ba^{2+}$ (142 pm) [34] Ca substitution should produce significant modifications in the structural topology of the CLBLCO compound (misfit strain between different layers and disorder effects within the layers). To address these features, we have performed high resolution powder x-ray diffraction (HRPXRD) study to evaluate the evolution of the structural disorder and lattice parameters as a function of oxygen and Ca contents.

Figure 7 shows a typical HRPXRD pattern of the $Ca_xLa_{1-x}Ba_{1.75-x}La_{0.25+x}Cu_3O_y$ compound together with its Rietveld analysis. The good Rietveld refinement of the diffraction data for all the samples and the absence of extra peaks further confirm the high quality and purity of our samples. The diffraction data were refined by using the FULLPROF program [37] by using a the pseudo-Voigt as peak-shape function and a polynomial function for the background fitting. The main outcome of the Rietveld refinement are reported in Table.2.

| x | y | a (Å) | c (Å) | $B_{Ca/La}$ (Å$^2$) | $B_{Ba/La}$ (Å$^2$) | $B_{Cu1}$ (Å$^2$) | $B_{Cu2}$ (Å$^2$) | $B_{O1}$ (Å$^2$) | $B_{O2}$ (Å$^2$) | $B_{O3}$ (Å$^2$) | $z_{Ba/La}$ | $z_{Cu2}$ | $z_{O2}$ | $z_{O3}$ | $\chi^2$ | Rwp |
|---|---|---|---|---|---|---|---|---|---|---|---|---|---|---|---|---|
| 0 | 7.1 | 3.9149(1) | 11.7604(1) | 0.35(5) | 0.97(5) | 1.04(5) | 0.42(5) | 3.8(5) | 0.55(8) | 1.2(2) | 0.1802(4) | 0.3451(4) | 0.366(1) | 0.157(1) | 3.6 | 11.8 |
| 0.1 | 7.1 | 3.9067(1) | 11.7480(1) | 0.36(5) | 1.04(5) | 1.22(5) | 0.46(5) | 5.6(5) | 0.61(8) | 1.3(2) | 0.176(1) | 0.346(1) | 0.360(1) | 0.154(1) | 2.6 | 10.2 |
| 0.2 | 7.1 | 3.8973(1) | 11.7304(1) | 0.29(5) | 1.16(5) | 1.29(5) | 0.52(5) | 7.5(5) | 0.54(8) | 1.6(2) | 0.1815(4) | 0.3475(4) | 0.366(1) | 0.155(1) | 2.6 | 10.3 |
| 0.3 | 7.1 | 3.8869(1) | 11.6985(1) | 0.27(5) | 1.19(5) | 1.38(5) | 0.58(5) | 5.9(5) | 0.56(8) | 1.6(2) | 0.1822(4) | 0.3479(4) | 0.366(1) | 0.160(1) | 3.3 | 12.0 |
| 0.3 | 6.9 | 3.8893(1) | 11.7203(1) | 0.28(5) | 1.25(5) | 1.50(5) | 0.59(5) | 5.0(5) | 0.51(8) | 2.0(2) | 0.1849(4) | 0.3490(4) | 0.365(1) | 0.157(1) | 3.5 | 11.9 |
| 0.3 | 6.4 | 3.8912(1) | 11.7971(1) | 0.23(5) | 1.17(5) | 2.00(5) | 0.57(5) | 3.6(5) | 0.35(8) | 2.3(2) | 0.1919(4) | 0.3511(4) | 0.367(1) | 0.155(1) | 5.6 | 20.1 |
| 0.4 | 7.1 | 3.8774(1) | 11.6685(1) | 0.29(5) | 1.23(5) | 1.55(5) | 0.59(5) | 5.3(5) | 0.59(8) | 1.8(2) | 0.1827(4) | 0.3489(4) | 0.366(1) | 0.160(1) | 5.6 | 17.3 |

**TABLE 2** structural parameters obtained by Rietveld refinement to the tetragonal P4/MMM structure: cell parameters a and c, isotropic atomic displacement factor $B_{iso}$ and z coordinates for different positions, $\chi^2$ and R factor.

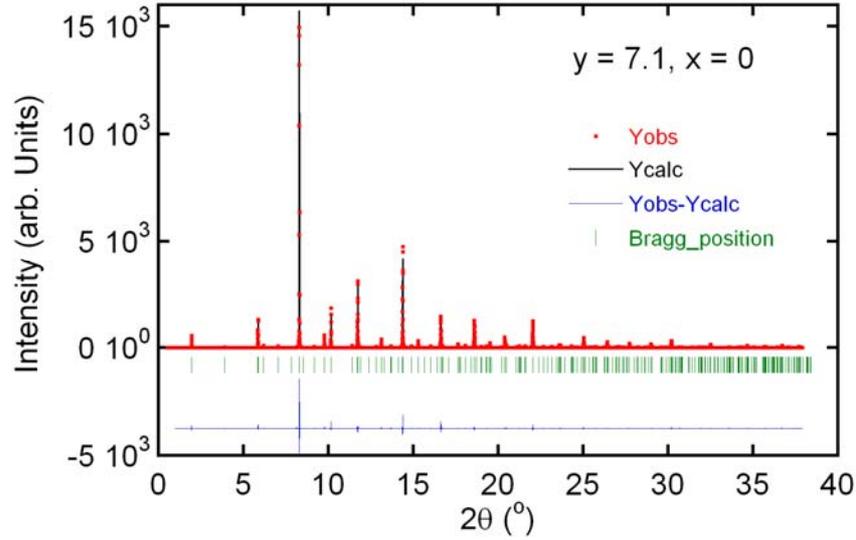

**Fig. 7** Rietveld refinement of a typical XRD pattern of a $Ca_xLa_{1-x}Ba_{1.75-x}La_{0.25+x}Cu_3O_y$ sample.

To evaluate the effect of changing Ca and oxygen contents on the structural disorder in CLBLCO, we have evaluated the widths of the two Bragg peaks (0 0 6) and (2 0 0) with different values of x and y (Fig. 8). The widths of the two peaks do not show any significant variation as a function of oxygen content (Fig. 9), which indicates that the average structural disorder is hardly affected by oxygen doping. Instead the broadening of both Bragg peaks for the sample with higher Ca content (Fig. 10) suggests that the introduction of Ca causes an increased disorder in CLBLCO. The Rietveld analysis reveals that the disorder is mainly located around the Ba site, indicated by the nearly doubling of the isotropic atomic displacement factor B (see Fig. 10). This result can be understood by considering that while the ionic radii of the $La^{3+}$ and $Ca^{2+}$ ions are quite similar, the ionic radius of the $Ba^{2+}$ ions is much bigger. Thus in CLBLCO the replacement of $La^{+3}$ ions for the $Ba^{+2}$ ions should cause a large local strain in the Ba layers. However this disorder cannot be responsible of the variation of $T_c^{max}$ as a function of Ca content. In fact the increased substitutional disorder should cause a reduction of $T_c$ [38] and not the remarkably increased $T_c^{max}$ from 45 K for x=0.0 to 81 K for x=0.4.

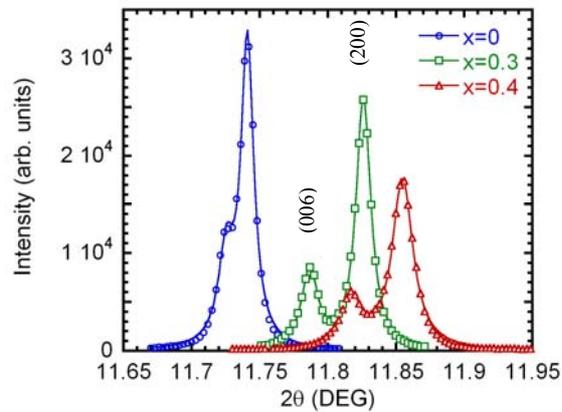

**Fig. 8** Representative diffraction profiles of the (006) and (200) Bragg peaks (symbols) with relative Lorentzian fits (solid curves) for the $Ca_xLa_{1-x}Ba_{1.75-x}La_{0.25+x}Cu_3O_y$ ($y \approx 7.1$).

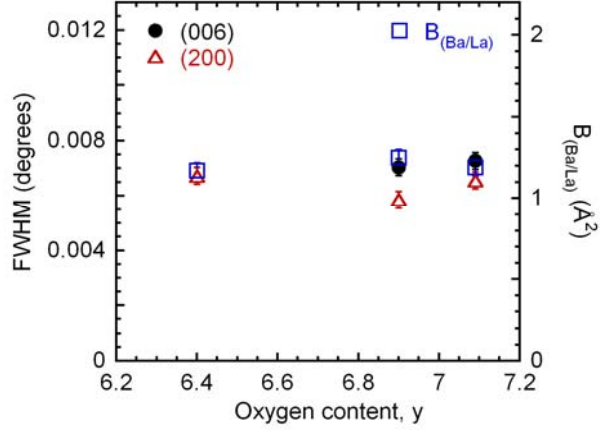

**Fig. 9** Line widths of the (006) and (200) Bragg peaks (left axis) and the isotropic atomic displacement factor B for the Ba/La site (right axis), obtained by the Rietveld analysis for the $Ca_xLa_{1-x}Ba_{1.75-x}La_{0.25+x}Cu_3O_y$ (x=0.3) as a function of the oxygen content.

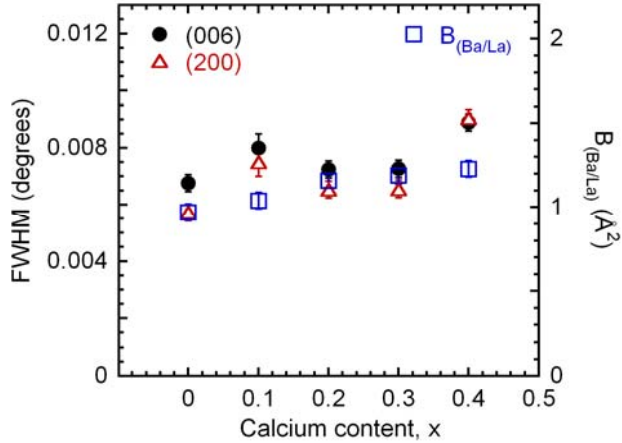

**Fig. 10** Line widths of the (006) and (200) Bragg peaks (left axis) and the isotropic atomic displacement factor B for the Ba/La site (right axis), obtained by the Rietveld analysis for the optimally doped $Ca_xLa_{1-x}Ba_{1.75-x}La_{0.25+x}Cu_3O_y$ as a function of Ca content

Intriguingly, the in-plane lattice parameter, which keeps constant upon O doping, displays a systematic linear relation with $T_c^{max}$ within the Ca content range 0≤x≤0.4 for the optimally doped CLBLCO samples (Fig. 11). The in-plane lattice parameter, a, closely relates to the Cu-O distance $R_{Cu-O}$, as $a = R_{Cu-O} \times (2\cos\theta)$, where θ is the buckling angle of the $CuO_2$ planes. Therefore, the increase of $T_c^{max}$ on decrease of inplane lattice parameter agrees with the more and more accepted mechanism that, in addition to the charge density, the misfit strain between different lattice parameters of different layers is another key parameter to control superconductivity in cuprates.

The intimate mechanism which links the misfit strain between different layers to superconductivity is still unclear. An elaborate analysis of the whole phase diagram of CLBLCO performed by Keren et al. [35] suggests that $T_c^{max}$ scales with the antiferromagnetic exchange interaction, hence providing support that the misfit strain control both magnetic interaction and superconductivity. However more recently J.L. Tallon et al. [3] have suggested that $T_c^{max}$ is driven by the polarizability of the charge-transfer layers, through either misfit strain and external pressure. Our results here indicate that CLBLCO is a very suitable compound to further investigate the role of polarizability in cuprates. In particular, studies on CLBLCO single crystals, whose growth has been achieved only recently [36], will permit to better investigate the intimate role of the structural changes on high temperature superconductivity.

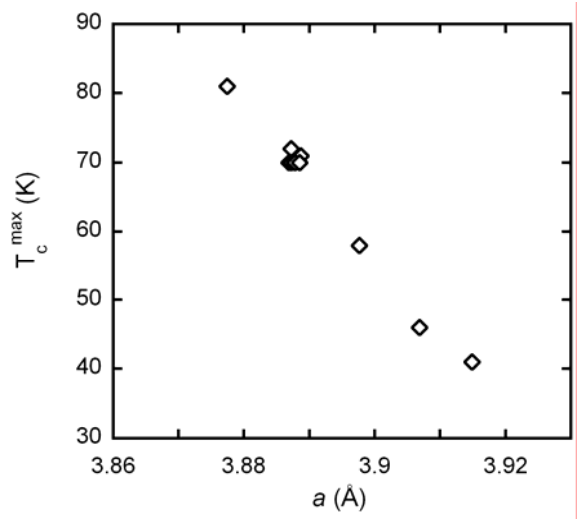

**Fig. 11** The linear relation between $T_c^{max}$ and inplane lattice parameter. The inplane lattice parameter, a, is obtained from the Rietveld analysis of the HRPXRD pattern of the optimum doped $Ca_xLa_{1-x}Ba_{1.75-x}La_{0.25+x}Cu_3O_y$ samples.

## 4. Summary

In summary, we have studied the electronic structure of unoccupied states in the charge-compensated $Ca_xLa_{1-x}Ba_{1.75-x}La_{0.25+x}Cu_3O_y$ (6.4≤y≤7.13, 0≤x≤0.4) system using O K-edge and Cu $L_3$-edge X-ray absorption spectroscopy. The analysis of XAS results provides information about the holes concentration and distribution in CLBLCO. Oxygen modifies superconductivity by doping holes simultaneously into apical sites, chains, and $CuO_2$ plane of CLBLCO while calcium substitution hardly changes the hole density or induces a charge transfer among different sites. Furthermore, the average disorder and lattice parameters were measured by high resolution powder X-ray diffraction. While average disorder appears to increase slightly with the Ca, around the Ba site, the in-plane lattice parameter reveals a systematic change with the superconducting transition temperature. On the basis of the combined analysis of electronic and lattice structures, we can conclude that $T_c^{max}$ in CLBLCO should be controlled by the misfit strain between different layers on the $CuO_2$ plane instead of the variation of charge density, charge transfer and/or substitutional disorder. Therefore, the work reported here provides direct evidence that there are other mechanisms in high temperature superconductivity besides hole doping, underlining the role of chemical pressure as an important driving parameter for superconductivity in CLBLCO cuprates. These results suggest the internal chemical pressure via misfit strain to be the most likely parameter to control the maximum critical temperatures ($T_c^{max}$) in different cuprate families at optimum hole density.

**Acknowledgements**

The authors gratefully acknowledge ESRF and BESSY for the beamtime and G. Mitdank and A. Ariffin for their assistance. We thank M. Filippi for help in HRPXRD measurements. This work was funded by PRIN06 "Search for critical parameters in High $T_c$ cuprates", the European Community-Research Infrastructure Action under the FP6 "Structuring the European Research Area" program (IA-SFS, Contract No. RII3-CT-2004-506008), as well as a partial support of the European project 517039 "Controlling Mesoscopic Phase Separation" (COMEPHS) (2005). We also acknowledge R. Ofer for providing the original data from Ref.[21], and A . Keren for fruitful discussion.